\documentclass{article}

\usepackage{arxiv}

\usepackage[utf8]{inputenc} 
\usepackage[T1]{fontenc}    
\usepackage{hyperref}       
\usepackage{url}            
\usepackage{booktabs}       
\usepackage{amsfonts}       
\usepackage{nicefrac}       
\usepackage{microtype}      
\usepackage{lipsum}		
\usepackage{graphicx}
\usepackage{natbib}
\usepackage{doi}
\usepackage{chemformula}
\usepackage{floatpag}
\usepackage{tabularx}
\usepackage{makecell}
\usepackage{soul}
\usepackage{multirow}
\usepackage[normalem]{ulem}
\usepackage[percent]{overpic}
\usepackage [autostyle, english = american]{csquotes}
\MakeOuterQuote{"}

\title{RADIATIVE FORCING BY \ch{CO_2} OBSERVED AT TOP OF ATMOSPHERE FROM 2002-2019}

\date{November 5, 2020}	

\author{ \href{https://orcid.org/0000-0001-7717-4744}{\includegraphics[scale=0.06]{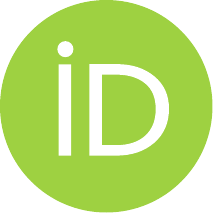}\hspace{1mm}Chris Rentsch}\thanks{\texttt{www.chrisrentsch.com}} \\
	Midland, MI 48642 \\
	\texttt{rentcp@gmail.com} \\
}

\renewcommand{\shorttitle}{\textit{Rentsch-2020-Radiative}}

\hypersetup{
pdftitle={RADIATIVE FORCING BY CO2 OBSERVED AT TOP OF ATMOSPHERE FROM 2002-2019},
pdfsubject={q-bio.NC, q-bio.QM},
pdfauthor={Chris Rentsch},
pdfkeywords={CO2, forcing, satellite, infrared, atmosphere, measurement},
}

\begin{document}
\maketitle

\begin{abstract}
	Spectroscopic measurements at top of atmosphere are uniquely capable of attributing changes in Earth's outgoing infrared radiation field to specific greenhouse gasses. 
	The Atmospheric Infrared Sounder (AIRS) placed in orbit in 2002 has spectroscopically resolved a portion of Earth's outgoing longwave radiation for over 17 years. 
	Concurrently, atmospheric \ch{CO_2} rose from 373 to 410 ppm, or 28\% of the total increase over pre-industrial levels. The Coupled Model Intercomparison Project Phase 6 (CMIP6) multi-model ensemble average predicts 0.477 Wm$^{-2}$ clear-sky longwave effective radiative forcing from this increase. Global measurements under nighttime, clear-sky conditions reveal 0.360$\pm$0.026 Wm$^{-2}$ of \ch{CO_2}-induced longwave radiative forcing, or 75$\pm$5\% of model predictions.
\end{abstract}


\section{Introduction}
\begin{figure}[h!]
		\centering
		\includegraphics[width=0.75\linewidth]{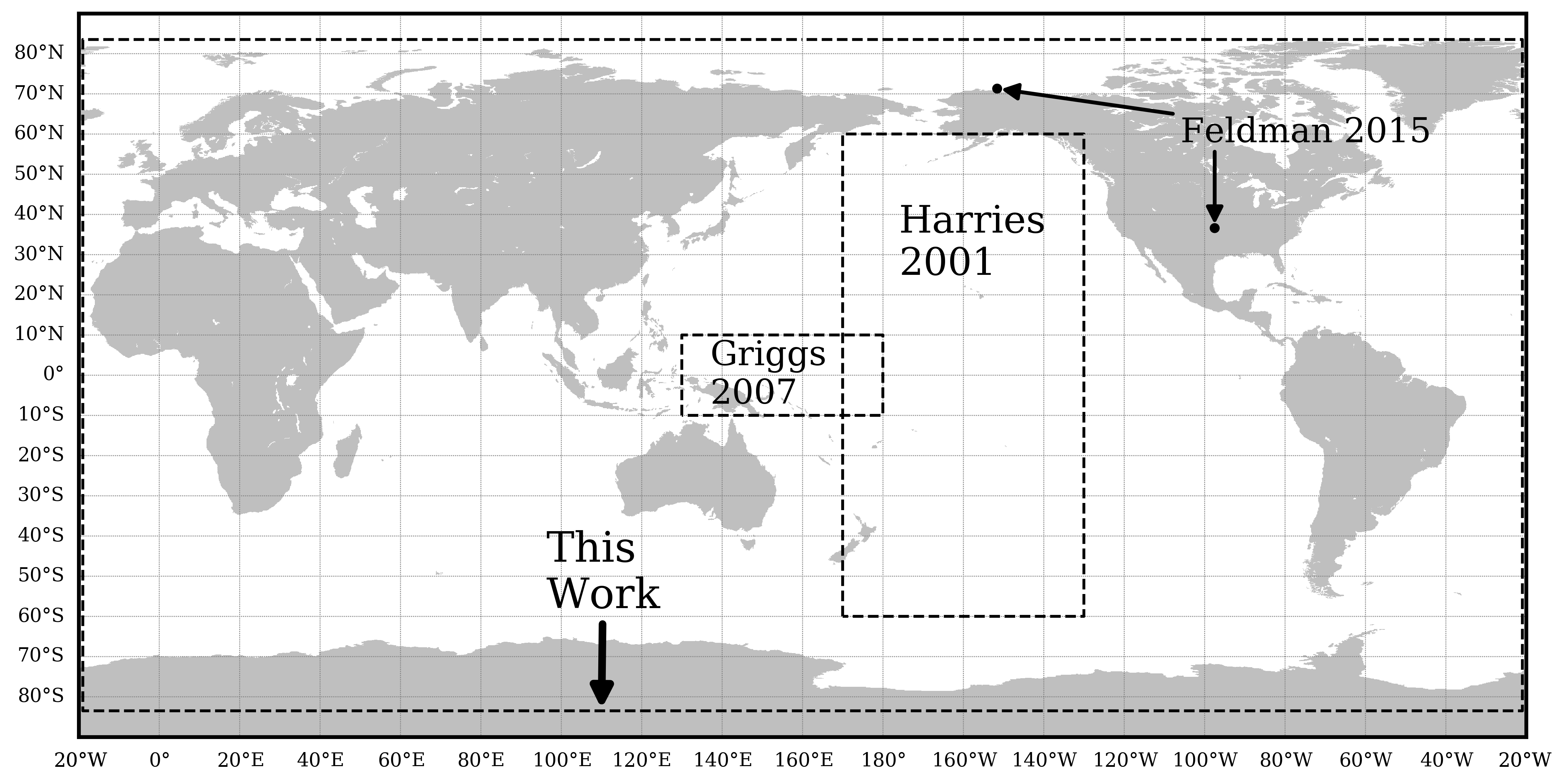}
		\caption{Measurement coverage for this work and select prior works by others}
		\label{fig:GlobalMappaper}
	\end{figure}

	Increasing infrared absorption caused by rising \ch{CO_2} is asserted as the fundamental physical mechanism underpinning anthropogenic global warming. Despite numerous studies of global temperature trends and rising greenhouse gas concentrations, very few investigations offer long-term spectroscopic measurement of \ch{CO_2} altering Earth's outgoing longwave radiation (OLR). \citet{harries2001increases} qualitatively compared 529 OLR spectra measured in 1970 by the IRIS satellite to 4,061 spectra measured in 1996 by IMG over the Pacific Ocean. \citet{griggs2007comparison} examined 25 IRIS, 138 IMG and 37,834 AIRS spectra measured over the central pacific. \citet{feldman2015observational} reported increasing downwelling longwave radiation (DLR) in two 1.6$^{\circ}$ conical upward views of the atmosphere between 2000 and 2010 (figure \ref{fig:GlobalMappaper}).
	
\begin{figure}
		\centering
		\includegraphics[width=0.75\linewidth]{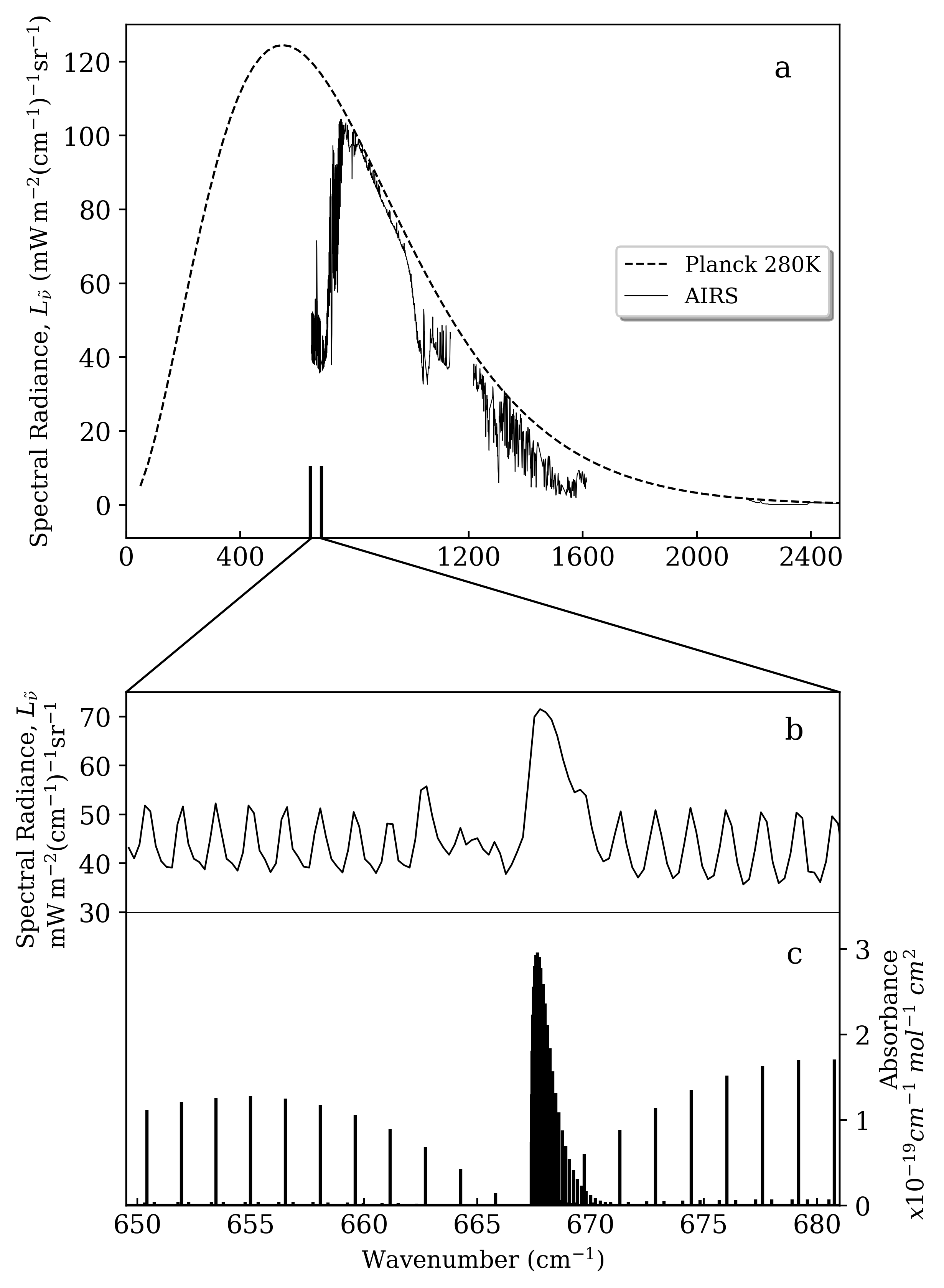}
		\caption{[a] AIRS nighttime, clear-sky OLR spectrum and 280K Planck distribution, [b] 650-680 cm$^{-1}$ subset, [c] HITRAN2016 spectral absorbance lines for \ch{CO_2} \citep{gordon2017hitran2016}. At wavenumbers where {\ch{CO_2}} is a strong IR absorber, it is also a strong emitter (Kirchhoff's Law) hence the excellent coincidence between detected radiance peaks and HITRAN \ch{CO_2} absorbance lines.}
		\label{fig:biplot}
	\end{figure}

	None provide a quantitative global assessment of \ch{CO_2}-induced radiative forcing. The Atmospheric Infrared Sounder (AIRS) offers the longest continuous record among all current or previous satellite spectrometers and has measured Earth's OLR while atmospheric \ch{CO_2} concentration rose from 373 to 410 ppm, 28\% of the total increase since 1750. Figure \ref{fig:biplot} exemplifies a single OLR spectrum comprised of 2,378 radiances. This work examines 42.4 billion global nighttime, clear-sky spectral radiance measurements (hereafter: radiances) made by AIRS between 2002-2019.  

\section{Data} 
\begin{figure}
	\centering
	\includegraphics[width=0.5\linewidth]{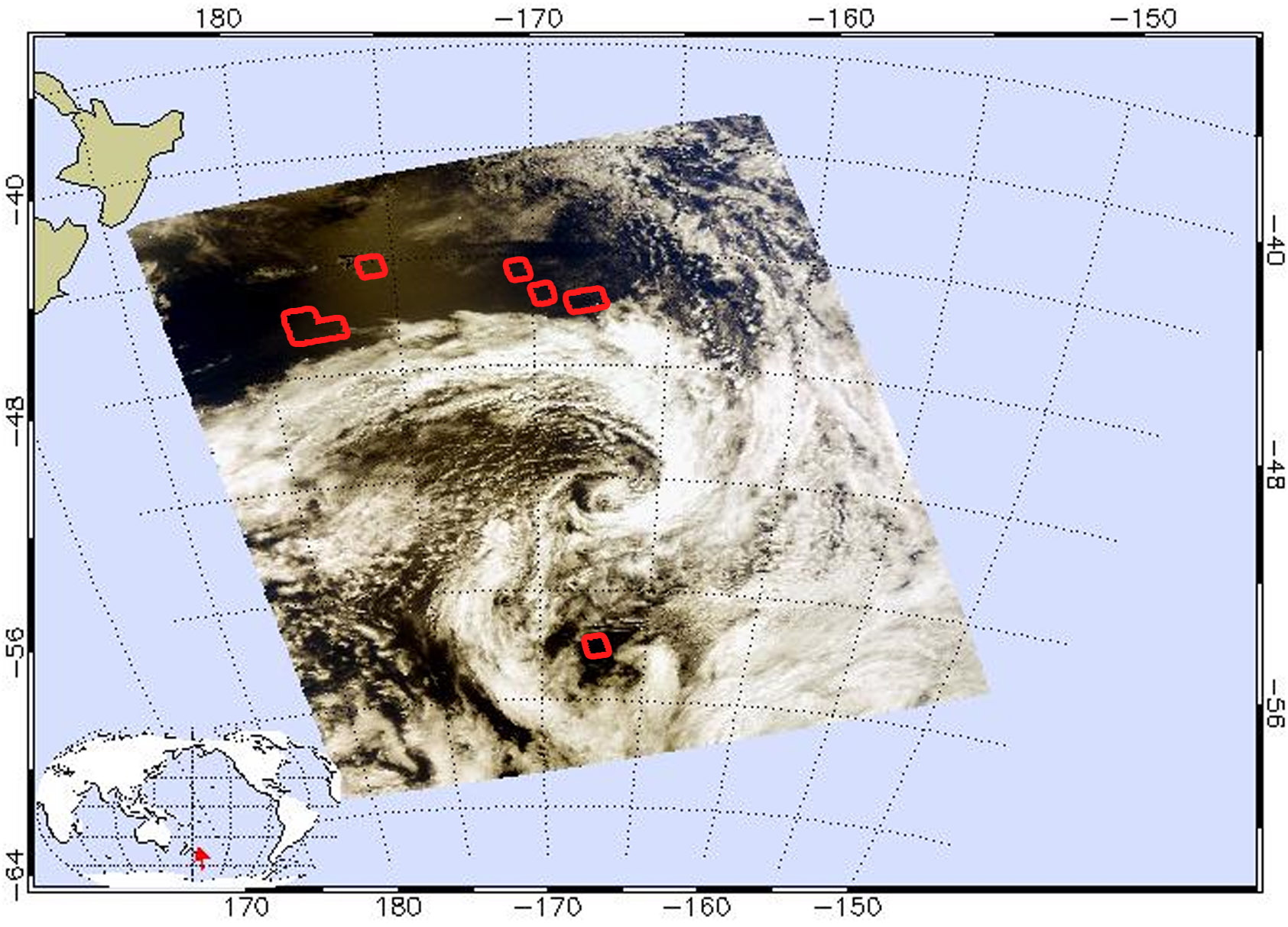}
	\caption{Visible wavelength image collected at 01:29:22 UTC on January 1, 2016. clear-sky regions identified by the cloud detection algorithm outlined in red.}
	\label{fig:cloudclearselection}
\end{figure}

	The majority of satellite views of Earth contain clouds that reflect or absorb upwelling infrared (IR). Clear-sky scenes are preferred to avoid attributing cloud-induced OLR reductions to \ch{CO_2}. The AIRS version 6 level 2 data product \citep{AIRS2CCF} quantifies fractional cloud content ranging from 0.00 to 1.00 for each radiance measurement. This work utilizes radiances with 0.00 cloud fraction; only 11\% of radiances meet this criterion. Although AIRS Level 2 measurements are a cloud-cleared data product, only naturally clear-sky radiances with no mathematical adjustments contribute to this analysis. Figure ~\ref{fig:cloudclearselection} provides a clear-sky selection example rendered from the sunlit side of Earth to permit comparison with a visible-wavelength image, however no daytime OLR measurements contributed to this analysis. It is evident that the cloud detection algorithm is conservative: visibly clear areas were not included and no visibly cloud-contaminated areas were inadvertently included (in this example the cloud detection algorithm has false positives but no false negatives).

	Solar longwave infrared radiation reflected by clouds or Earth's surface and combining with terrestrial IR would contaminate daytime OLR measurements. To eliminate this source of error, only nighttime measurements were utilized. When the AIRS solar zenith angle (SZA) is $\leq$90$^{\circ}$ AIRS observes the sunlit side of Earth, when 90$^{\circ}$$<$SZA$<$108$^{\circ}$ it observes the twilight region and when SZA$\geq$108$^{\circ}$ it observes the nighttime region. Only measurements at SZA$\geq$108$^{\circ}$ were utilized.  

	The AIRS mirror scans $\pm$49.5$^{\circ}$ from nadir across flight track. In the bands of radiatively active gasses, higher scan angle observations exhibit limb-brightening or limb-darkening due to IR emission originating higher in the atmosphere. To reduce the effects of this anisotropy, only radiances at scan angles $\leq$15$^{\circ}$ were utilized.

	AIRS does not have measurement capability at $<$649.6 cm$^{-1}$, 1136-1217 cm$^{-1}$ or 1614-2181 cm$^{-1}$. The 2,378 individual channels comprising the AIRS IR sensor are monitored for quality and only those recommended for scientific use and not subject to static or dynamic noise were utilized. Radiance measurements flagged as dust-contaminated were excluded, though these were rare ($<$0.01\%). 

\section{Method}
\begin{figure}
	\centering
	\includegraphics[width=0.8\linewidth]{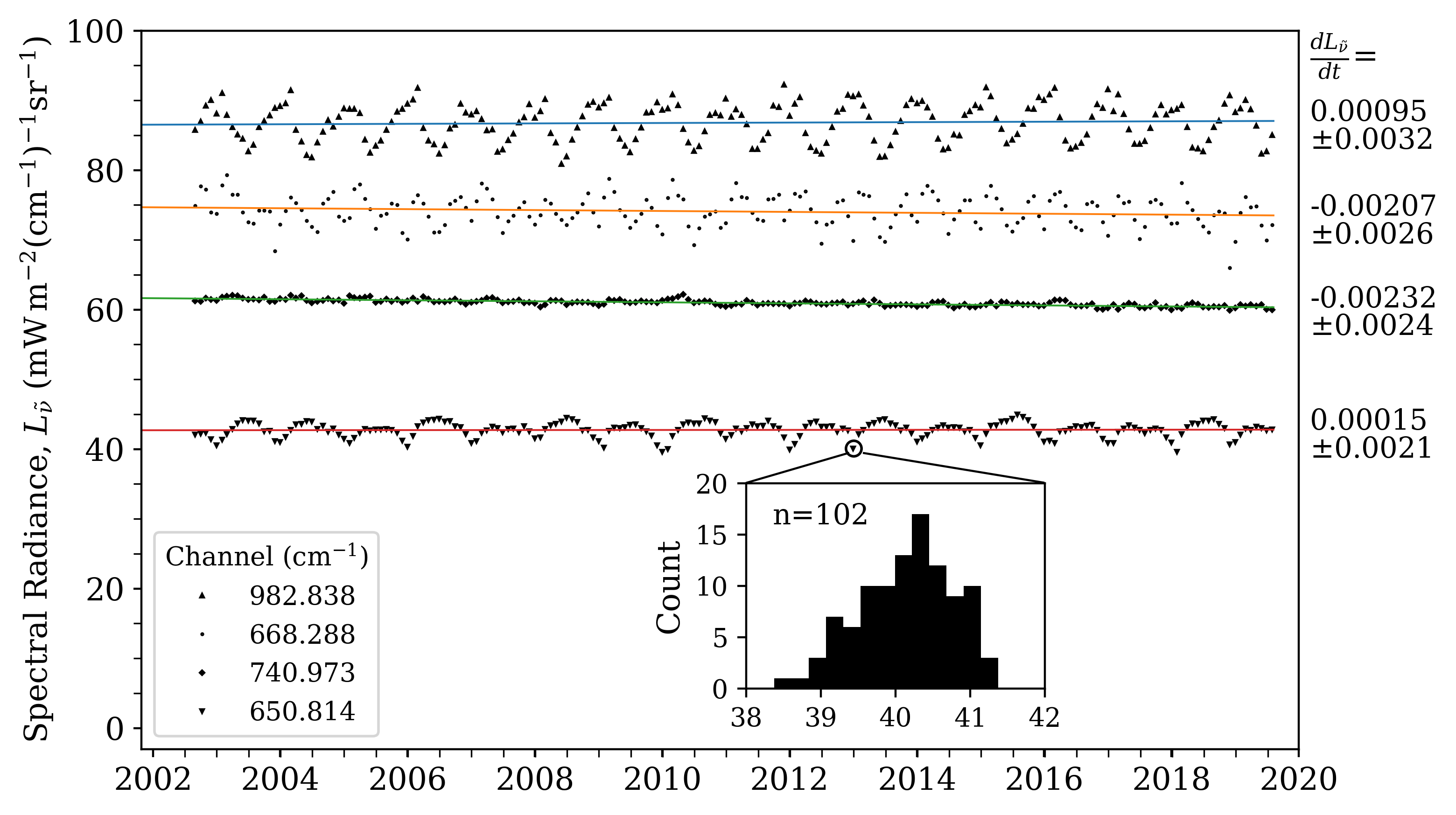}
	\caption[RegExample]{Least-squares regression fit to average monthly radiances for four AIRS channels in 0$^{\circ}$-10$^{\circ}$S x 20$^{\circ}$-40$^{\circ}$E grid cell. Remaining channels in all 324 grid cells were fit similarly.}
	\label{fig:RegExample}
\end{figure}

	Every AIRS radiance measurement meeting the selection criteria was analyzed in this study (not a subset). The nature of the satellite's precessing polar orbit characteristically produces fewer equatorial observations and more polar observations than area-proportional to Earth. To generate a spatially-representative dataset, radiances for a given channel in a given month were gridded in 10$^{\circ}$ latitude x 20$^{\circ}$ longitude cells. For example, 102 nighttime, clear-sky radiances measured at 0$^{\circ}$-10$^{\circ}$S x 20$^{\circ}$-40$^{\circ}$E contribute to the 650.814 cm$^{-1}$ average radiance for January 2013 (figure ~\ref{fig:RegExample}, inset). 25.1\% of grid cells contain no data due to heavy clouds, lack of nighttime measurements (e.g., polar summers) or failed detector channels. A median of 65 radiances contribute the monthly average for a given channel in a given grid cell. Over time, some detector channels succumb to solar radiation exposure and cease useful data production. Of the maximum potential 17 year record, channels with fewer than four years were excluded for insufficient record length (0.21\% of all channels).  

\begin{figure*}
	\centering
	\includegraphics[width=1.05\textwidth]{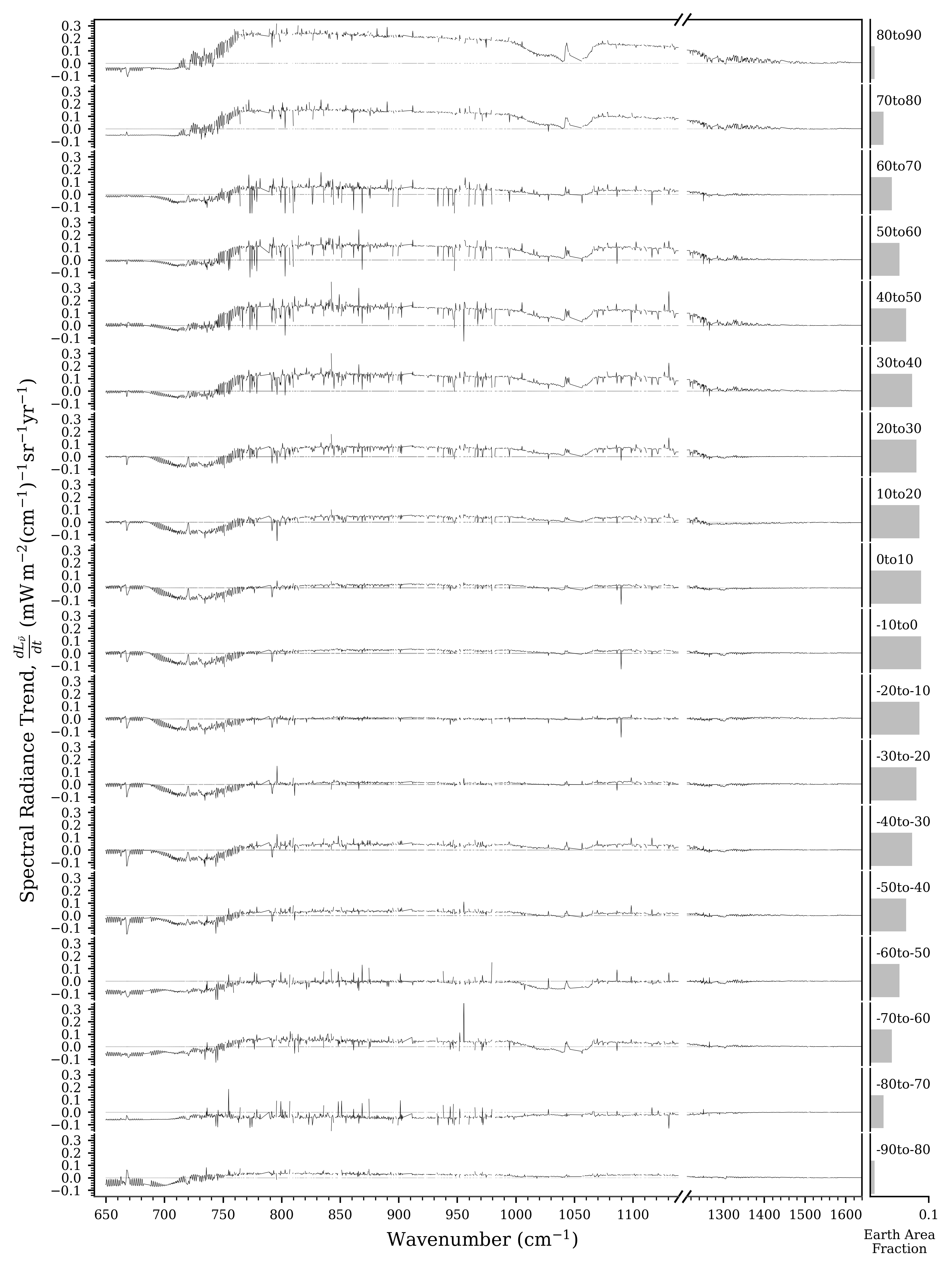}
	\caption{Nighttime, clear-sky spectral radiance trend 2002-2019 in 10$^{\circ}$ latitude increments. $\pm$ 2.3 mK/yr shaded at each abscissa.}
	\label{fig:Trends_by_latitude}
\end{figure*}

\begin{figure*}[h!]
	\centering
	\includegraphics[width=1.0\textwidth]{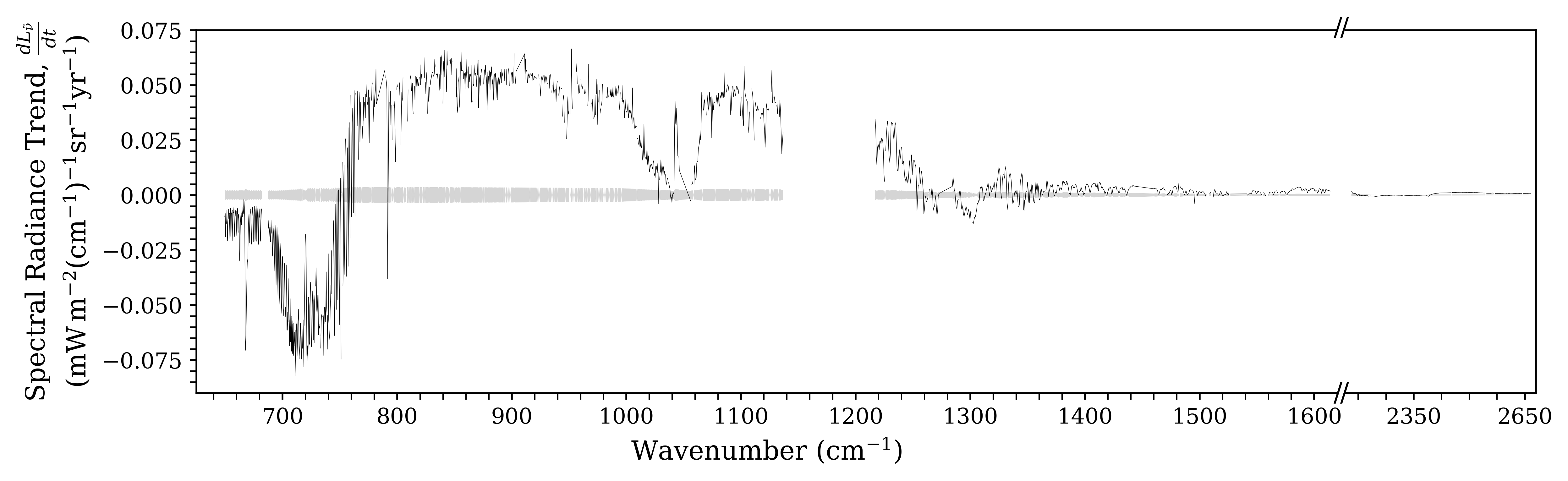}
	\caption{Global composite nighttime, clear-sky spectral radiance trend 2002-2019. $\pm$2.3 mK/yr shaded at $<$1614 cm$^{-1}$ and $\pm$8 mK/yr shaded at $>$2181 cm$^{-1}$. OLR reductions attributable to +37 ppm \ch{CO_2} are evident at $<$765 cm$^{-1}$. Increased emission in the atmospheric window 800-1000 cm$^{-1}$ is consistent with $\approx$0.3$^{\circ}$C/decade surface warming.}
	\label{fig:trend}
\end{figure*}

	A straight line was fit by least-squares regression to the time series of monthly radiance averages for each channel in each grid cell. Seasonal temperature cycles were eliminated as a source of trend bias by utilizing only complete years of measurement data starting on 1 September 2002 and ending on 31 August 2019. An example line fitting is provided in figure ~\ref{fig:RegExample}. The slope of each line is the spectral radiance trend $\frac{dL_{\tilde{\nu}}}{dt}$ (mW$\,$m$^{-2}$(cm$^{-1}$)$^{-1}$sr$^{-1}$yr$^{-1}$) and the trend uncertainty is $\pm 2.3$mK/yr, the AIRS instrument stability reported by \citet{strow2020establishment}. Instrument stability was used instead of the standard error of the regression because the latter is overwhelmingly an indication of seasonality in the radiance time series. A larger uncertainty of $\pm8$mK/yr was utilized at midwave IR channels (2181-2665 cm$^{-1}$) for reasons detailed in sources of error. Lines were fit to monthly averages of all channels in all 324 grid cells. Area-weighted spectral radiance trends for 10$^{\circ}$ latitude increments (figure~{\ref{fig:Trends_by_latitude}}) and the global average (figure~{\ref{fig:trend}}) were constructed therefrom. The scan angle restriction prevents measurement at $>$83.5$^{\circ}$N and $<$83.5$^{\circ}$S therefore polar radiances measured at 80$^{\circ}$-83.5$^{\circ}$ were weighted for 80$^{\circ}$-90$^{\circ}$.

\section{Results: \ch{CO_2} Radiative Forcing}
	OLR reductions in the \ch{CO_2} $v_2$ and $v_3$ bands are presumed the result of rising atmospheric \ch{CO_2} concentration. The reductions at detectable portions of the $v_2$ band fundamental Q-branch (667 cm$^{-1}$) and associated P- and R-branches (650-687 cm$^{-1}$) are relatively minor compared to 687-765 cm$^{-1}$ where the majority of detectable OLR reduction coincides with the 720 cm$^{-1}$ Q-branch and associated P- and R-branches, frequently described as a wing. Another Q-branch at 791 cm$^{-1}$ was also quantified, though its P- and R-branches do not absorb strongly enough to reduce OLR at current \ch{CO_2} concentrations. The $v_3$ asymmetric stretch generates a parallel band with P- and R-branches but no central Q-branch. The \ch{C^{12}O_2} and \ch{C^{13}O_2} $v_3$ bands are centered at 2349 cm$^{-1}$ and 2284 cm$^{-1}$, respectively. 

\begin{figure}
	\centering
	\includegraphics[width=1.0\linewidth]{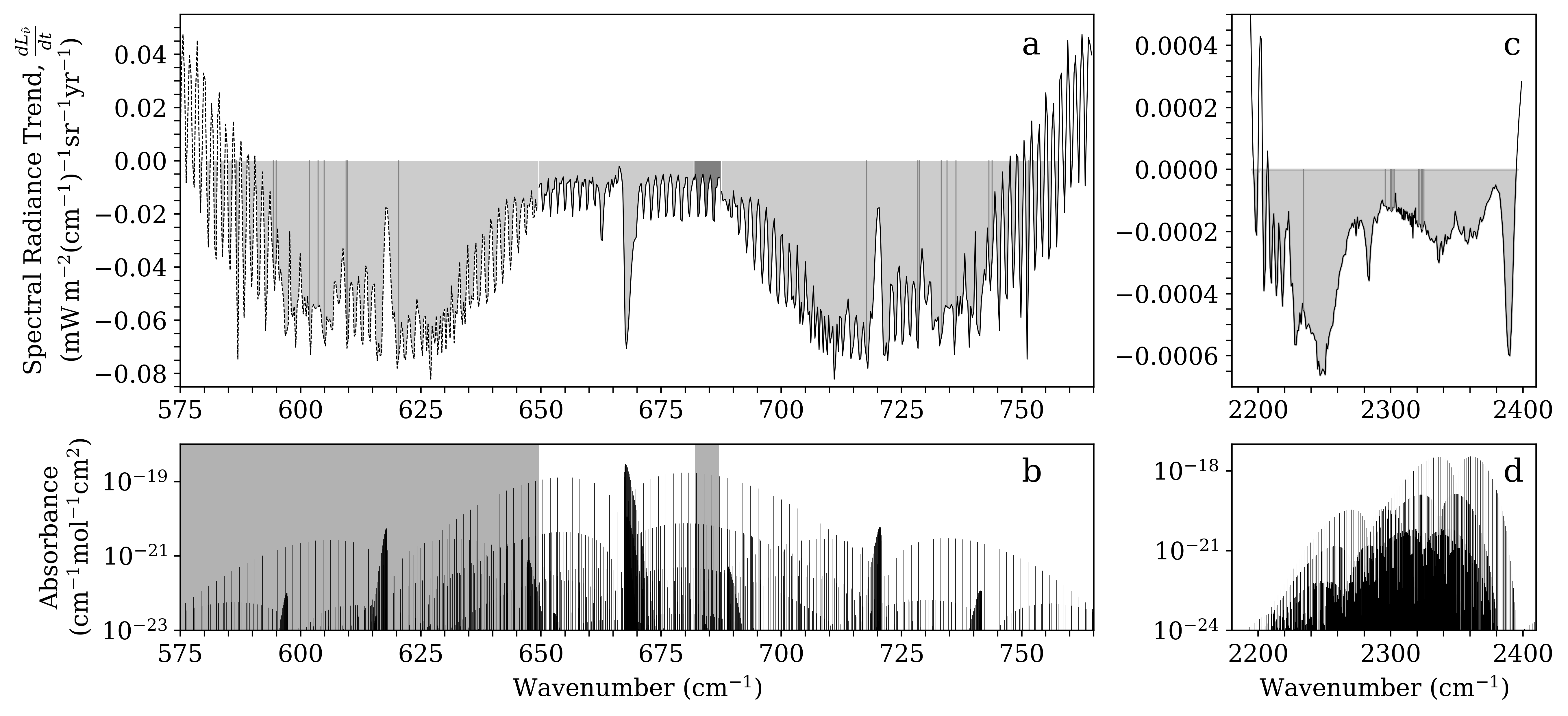}
	\caption{[a] {\ch{CO_2}} $v_2$ band spectral radiance trends. Integrated regions shaded: measured values in light gray and interpolations in dark gray. Dashed line at 575-649 cm$^{-1}$ is 687-765 cm$^{-1}$ measurements reflected about the 667 cm$^{-1}$ Q-branch. [b] HITRAN 2016 absorption lines for $v_2$ band. Shaded regions are where AIRS lacks measurement capability. [c] and [d] are the same as [a] and [b] except for the $v_3$ band.}
	\label{fig:integrals}
\end{figure}

	OLR radiant exitance change $\delta M$ (Wm$^{-2}$) was produced by integrating the spectral radiance trend ($\frac{dL_{\tilde{\nu}}}{dt}$) over the range of \ch{CO_2}-affected wavenumbers and multiplying by 17 years and by $\pi\,$sr, regarding the atmosphere as a Lambertian emitter at these optically-thick channels:
	\begin{equation} \label{eq:1}
		\delta M_{v_2PQR}=17\pi\int_{649.6 cm^{-1}}^{681.7 cm^{-1}} \frac{dL_{\tilde{\nu}}}{dt} d\tilde{\nu} = -0.029\pm 0.004 \, Wm^{-2}
	\end{equation}
	\begin{equation} \label{eq:2}
		\delta M_{v_2wing}=17\pi\int_{687.6 cm^{-1}}^{764.5 cm^{-1}} \frac{dL_{\tilde{\nu}}}{dt} d\tilde{\nu} = -0.164\pm 0.011 \, Wm^{-2}
	\end{equation}
	\begin{equation} \label{eq:3}
		\delta M_{v_2 791}=17\pi\int_{791.4 cm^{-1}}^{792.5 cm^{-1}} \frac{dL_{\tilde{\nu}}}{dt} d\tilde{\nu} = -0.0008\pm 0.0003 \, Wm^{-2}
	\end{equation}
	\begin{equation} \label{eq:4}
		\delta M_{v_3}=17\pi\int_{2195 cm^{-1}}^{2396 cm^{-1}} \frac{dL_{\tilde{\nu}}}{dt} d\tilde{\nu} = -0.003\pm 0.001 \, Wm^{-2}
	\end{equation}
	Integrals are depicted in figure \ref{fig:integrals}. At the wing edges in (\ref{eq:2}) and (\ref{eq:4}), only reductions in OLR were integrated. A measurement gap between AIRS detector arrays 11 and 12 (681.7-687.6 cm$^{-1}$) was estimated using 676.6-681.7 cm$^{-1}$ R-branch measurements. Additionally, fourteen individual failed detector channels were interpolated from adjacent channels prior to integration. The majority of detectable OLR change is attributable to the increasing \ch{CO_2} $v_2$ wing absorption at 687-765 cm$^{-1}$. The symmetrical wing at 575-650 cm$^{-1}$ is outside of AIRS measurement range.
	Consequently, the sum of (\ref{eq:1}-\ref{eq:4}) is only a partial measurement of $\delta$M caused by rising \ch{CO_2} and total $\delta M_{LW}$ must be estimated. The P-branch and R-branch absorption lines flanking the Q-branch are nearly symmetrical and rising \ch{CO_2} caused nearly-identical reductions in radiance. By extension, it is a reasonable prediction that the unmeasured 575-650 cm$^{-1}$ wing has undergone OLR reduction by an amount similar to the measured 687-765 cm$^{-1}$ wing. The presumption of wing symmetry is supported by several radiative transfer model simulations of top-of-atmosphere (TOA) OLR spectral change after doubling atmospheric \ch{CO_2} \citep{charlock1984co2, huang2009evolution, huang2010separation, brindley2016spectral}. Therefore, total OLR reduction due to rising \ch{CO_2} is estimated as:
	\begin{equation} \label{eq:5}
		\delta M_{LW}=\delta M_{v_2PQR} + 2 \delta M_{v_2wing} + \delta M_{v_2 791} + \delta M_{v_3} = -0.360\pm 0.026 \, Wm^{-2}
	\end{equation}
	The uncertainty is taken as the simple sum of the component uncertainties rather than by quadrature as causes of instrument drift are unlikely to be independent and are rather more likely related to a common optical system component (e.g., black body cavity).
\section{Comparison to CMIP6}		
	Measurement of the atmospheric \ch{CO_2} concentration increase that produced $\delta M_{LW}$ was supplied by the NOAA ESRL's global monitoring division \citep{NOAA}. The combination of empirical measurements of RF at TOA and \ch{CO_2} concentration change permits comparison to climate model predictions of \ch{CO_2}-induced radiative forcing. AIRS TOA measurements contain all infrared responses to surface and atmospheric temperature adjustments, therefore the most suitable comparison is effective radiative forcing (ERF), introduced by \citet{myhre2013anthropogenic}. ERFs calculated by high spectral resolution line-by-line models at fixed surface temperature give no adjustment for increased OLR from the Planck response to surface warming. More computationally-intensive climate models are required to generate ERFs that include surface temperature change, with the trade-off of utilizing a lower spectral resolution \citep{etminan2016radiative}. Most climate models in the Coupled Model Intercomparison Project Phase 5 (CMIP5) do not fully resolve the stratosphere \citep{flato2014evaluation}. The majority of the CMIP6 models examined by \citet{smith2020effective} resolve the atmosphere to $>$80km and all compute a stratospheric temperature adjustment. The CMIP6 multi-model ensemble is therefore a more suitable comparison to AIRS measurements. Clear-sky longwave {\ch{CO_2}} ERFs were produced by taking the difference between TOA outgoing clear-sky longwave radiation for CMIP6 4 x {\ch{CO_2}} and Control model runs and averaging the years after sufficient time elapsed to homogenize atmospheric {\ch{CO_2}} concentration, typically 1 year \citep{smith2020effective}. The 4 x {\ch{CO_2}} ERFs were scaled down to 1.1 x {\ch{CO_2}} using the forcing relationship from \citet{etminan2016radiative} to accord with the atmospheric concentration increase from 2002-2019. The resulting 1.1 x {\ch{CO_2}} ERFs are compared to AIRS measurement in table \ref{tab:ForcingResultsTable}. All CMIP6 climate models predict 0.07-0.16 Wm$^{-2}$ more longwave forcing than is observed. Potential causes of model-observation disagreement are beyond the scope of this work, but it is notable that difference between AIRS observation and the model giving the lowest forcing is approximately the same magnitude as difference between the models giving the highest and lowest forcing.

\begin{table}[h]
	\centering
	\caption {Longwave clear-sky effective radiative forcing from CMIP6 multi-model ensemble scaled down to 1.1 x \ch{CO_2} to compare to AIRS 2002-2019 measurements (373.13$\rightarrow$410.21 ppm).}
	\label{tab:ForcingResultsTable}
	\begin{tabular}{@{}llr@{}}
		\toprule
		Source & Wm$^{-2}$ & vs AIRS\\ 
		\midrule
		ACCESS-CM2 \citep{https://doi.org/10.22033/ESGF/CMIP6.4283} & {0.471} & {+31}\%\\
		CanESM5	\citep{https://doi.org/10.22033/ESGF/CMIP6.3659} & {0.451} & {+25}\%\\
		CESM2 \citep{https://doi.org/10.22033/ESGF/CMIP6.7706} & {0.499}	& {+38}\%\\
		CNRM-CM6-1 \citep{https://doi.org/10.22033/ESGF/CMIP6.4136}	& {0.459} & {+27}\%\\
		CNRM-ESM2-1 \citep{https://doi.org/10.22033/ESGF/CMIP6.9643}	& {0.467} & {+30}\%\\
		EC-Earth3 \citep{https://doi.org/10.22033/ESGF/CMIP6.4815} & {0.502}	& {+39}\%\\
		GFDL-CM4 \citep{https://doi.org/10.22033/ESGF/CMIP6.8638} & {0.486}	& {+35}\%\\
		GFDL-ESM4 \citep{https://doi.org/10.22033/ESGF/CMIP6.11964} & {0.453} & {+26}\%\\
		GISS-E2-1-G	\citep{https://doi.org/10.22033/ESGF/CMIP6.7312} & {0.495} & {+37}\%\\
		HadGEM3-GC31-LL	\citep{https://doi.org/10.22033/ESGF/CMIP6.6224} & {0.460} & {+28}\%\\
		IPSL-CM6A-LR \citep{https://doi.org/10.22033/ESGF/CMIP6.5236} & {0.463} & {+28}\%\\
		MIROC6 \citep{https://doi.org/10.22033/ESGF/CMIP6.5682} & {0.497} & {+38}\%\\
		MPI-ESM1-2-LR \citep{https://doi.org/10.22033/ESGF/CMIP6.6659} & {0.509} & {+41}\%\\
		MRI-ESM2-0	\citep{https://doi.org/10.22033/ESGF/CMIP6.6873} & {0.444} & {+23}\%\\
		NorESM2-LM \citep{https://doi.org/10.22033/ESGF/CMIP6.8122} & {0.516} & {+43}\%\\
		NorESM2-MM \citep{https://doi.org/10.22033/ESGF/CMIP6.8124} & {0.510} & {+42}\%\\
		UKESM1-0-LL \citep{https://doi.org/10.22033/ESGF/CMIP6.11086} & {0.431} & {+20}\%\\
		\midrule
		CMIP6 multi-model ensemble average & {0.477} & {+32}\%\\
		AIRS -$\delta M_{LW}$ & 0.360$\pm$0.026 & ---\% \\ 
		\bottomrule
	\end{tabular}
\end{table}

\section{Sources of Error}
	The unmeasured {\ch{CO_2} $v_2$ wing was assumed to undergo radiance reduction identical to the measured wing, however, 575-650 cm$^{-1}$ overlaps with stronger water vapor absorption lines. The assumption of symmetry is conservatively high and actual OLR reduction at 575-650 cm$^{-1}$ is expected to be slightly less than at 687-765 cm$^{-1}$. Trend asymmetry between the two wings was observed in DLR reported by \citet{feldman2015observational}: the 575-650 cm$^{-1}$ wing showed less forcing change over time relative to the 687-765 cm$^{-1}$ wing, particularly in the southern great plains where atmospheric moisture content is higher. 
	
	Over 17 years, detector stability is more important than absolute accuracy as unbiased noise averages to zero over large statistical samples. One possible cause of systematic trend bias is gradual accumulation of molecular contaminants on the AIRS detector mirror. A hypothetical 100\AA~contamination layer is predicted to increase the mirror emissivity variation by 0.001, producing cold scene brightness temperatures at 650-800 cm$^{-1}$ that are 0.1-0.2$^{\circ}$ K warmer than reality \citep{aumann2000airs}. If such a contamination layer were gradually building up during the observation period, warming trends could be amplified and cooling trends (including forcing) could be diminished. \citet{aumann2018radiometric} reported evidence of mirror contamination between 2002-2010 affecting AIRS midwave IR channels (2181-2665 cm$^{-1}$) which were utilized to compute the \ch{CO_2} $v_3$ forcing, though this band contributes $<$1\% to total observed forcing.
	\begin{figure}[h]
		\centering
		\includegraphics[width=0.75\linewidth]{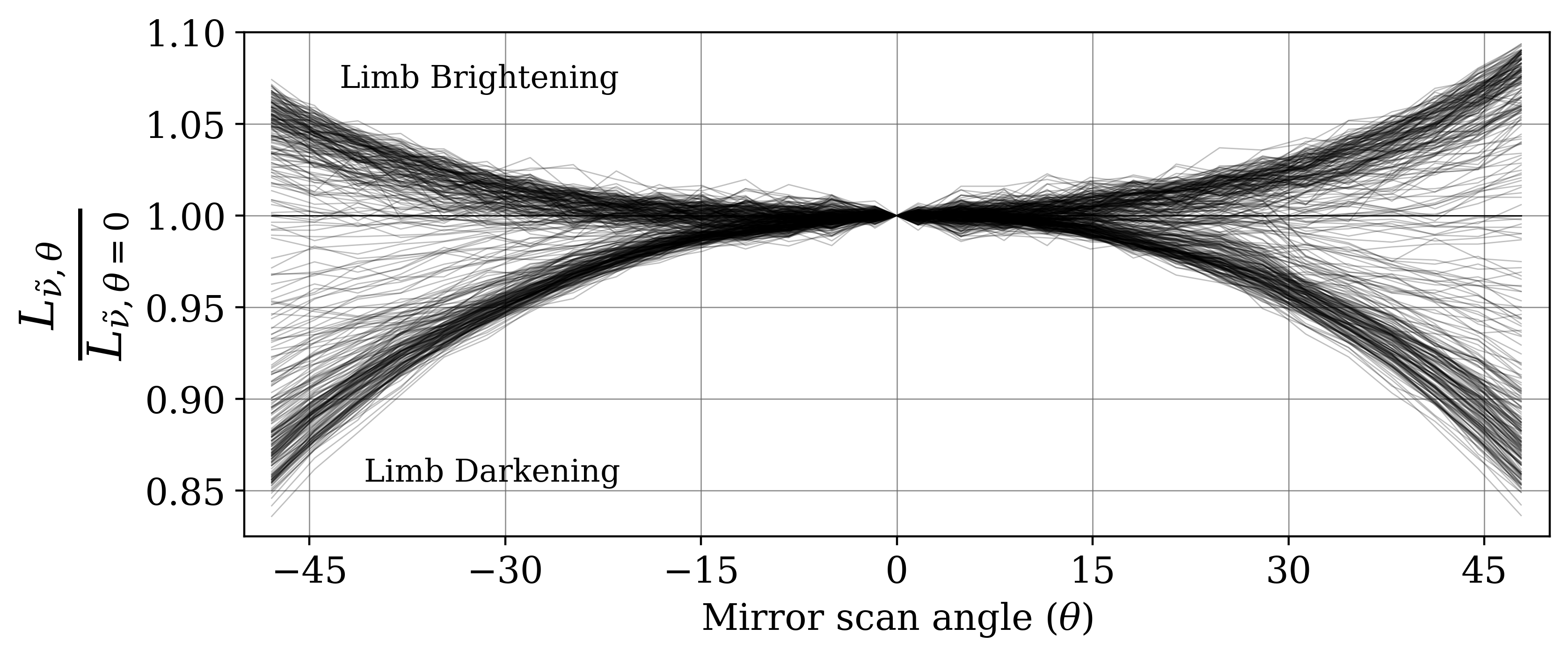}
		\caption{Limb brightening/darkening for 650-750 cm$^{-1}$ from granule 004 on Jan 5, 2016. Only radiances at $|\theta| \leq 15 ^{\circ}$ were utilized.}
		\label{fig:Limbs}
	\end{figure}
	
	This analysis assumed isotropic atmospheric emission when computing radiant exitance change from spectral radiance change. However, high scan angle measurements rarely satisfy the condition of isotropy. Slant path views through an optically-thick atmosphere are subject to significant limb-brightening or limb-darkening. The variation of spectral radiance with scan angle ($\theta$) is apparent upon examining the ratio of $L_{\tilde{\nu},\theta}$ to the near-simultaneous nadir measurement $L_{\tilde{\nu},\theta =0}$ at the same wavenumber (figure \ref{fig:Limbs}). Over time, if proportionally more (or fewer) high-angle measurements meet the quality and clear-sky selection criteria, trend bias would result. Others have addressed this by applying a computationally-expensive direct integration \citep{doniki2015instantaneous} or by excluding higher scan angle measurements \citep{aumann2006three, griggs2007comparison}. Given the abundance of data available to this study, the latter approach was adopted and radiances most affected by limb-brightening/darkening at $|\theta| >$15$^{\circ}$ were excluded.
	
\section{Conclusion}
	Climate models of the Earth energy system may contain errors of opposite signs in different spectral bands that fortuitously compensate, providing satisfactory model agreement for nonphysical reasons \citep{huang2007strict, etminan2016radiative}. Empirical measurement by hyperspectral satellites provides a means to adjudicate between different climate model RF calculations and promote only those matching reality. Seventeen years of AIRS nighttime, clear-sky OLR measurements reveal 0.360$\pm$0.026 Wm$^{-2}$ additional longwave radiative forcing induced by +37 ppm atmospheric \ch{CO_2}. AIRS lacks measurement capability at 575-650 cm$^{-1}$ for complete \ch{CO_2} $v_2$ band characterization, therefore this empirical estimate of increased forcing was devised by presuming \ch{CO_2} $v_2$ wing symmetry and doubling the observed wing's radiative forcing. The latest generation CMIP6 climate models predict 0.431-0.516 Wm$^{-2}$ of clear-sky longwave ERF, +20-43\% greater than observed by AIRS. Current climate models may require modest revision to bring {\ch{CO_2}} forcing computations into agreement with observation.







\end{document}